\documentclass[preprint,journal]{vgtc}       % preprint (journal style)
\usepackage{mathptmx}
\usepackage{graphicx}
\usepackage{epstopdf}
\usepackage{times}
\usepackage{url}
\onlineid{0}

\vgtccategory{Research}

\vgtcinsertpkg

\title{Design Decisions for Weave: A Real-Time Web-based Collaborative Visualization Framework}

\author{Andrew Dufilie and Georges Grinstein} \authorfooter{
 Andrew Dufilie and Georges Grinstein are with the Institute for Visualization and Perception
 Research at the University of Massachusetts at Lowell. adufilie@cs.uml.edu, grinstein@uml.edu.
}

\shortauthortitle{Dufilie \MakeLowercase{\textit{et al.}}: Design Decisions for Weave}
\abstract{There are many web-based visualization systems available to date, each
having its strengths and limitations. The goals these systems set out to
accomplish influence design decisions and determine how reusable and scalable
they are.  Weave is a new web-based visualization platform with the broad goal
of enabling visualization of any available data by anyone for any purpose. Our
open source framework supports highly interactive linked visualizations for
users of varying skill levels.  What sets Weave apart from other systems is its
consideration for real-time remote collaboration with session history.  We
provide a detailed account of the various framework designs we considered with
comparisons to existing state-of-the-art systems. }

\begin{document}


\firstsection{Introduction}

\maketitle

%% \section{Introduction} %for journal use above \firstsection{..} instead

There are many different ways of implementing online visualization systems.
Plots can be rendered by a server, a desktop client, or a web client. One of the
fastest ways to display plots of large amounts of data in a web-based
visualization is to use images generated by a server.  This approach is used by
Tableau public.\cite{tableau}  The images can be transferred much faster than
the data used to generate them, so the user receives quick initial view of the
data, which makes a good first impression.  An additional advantage is that a
server-side implementation allows computationally expensive analysis algorithms
to be applied to data.  However, depending on the level of interaction and type
of visual feedback desired, server-side rendering may not be desirable.

In server-side visualization systems where the client does not have the actual
data used to generate the plots, responsive interactions may become an issue
depending on the speed of the server.  If the server does not respond quickly
enough, the system may seem sluggish and awkward, making the user feel distant
from the visualization system.  For this reason, we prefer a client-side
implementation.

The Weave project was launched in 2008, with ActionScript 3.0 as the language of
choice because of the Flash Player's graphics and interaction capabilities as
well as its worldwide ubiquity\cite{ubiquity}, which meant Weave would be
immediately accessible to a wide market without the need for installing a new
piece of software.  

The evolution of Weave has been supported and guided by members of the Open
Indicators Consortium (OIC)\cite{oic}.  The general goal of the Weave project is
to create a scalable web-based platform that supports data visualization,
exploration, analysis, session history, and real-time collaboration.  Baumann et
al.\cite{baumann} provide an in-depth discussion about the design goals of Weave
related to session history and collaboration.

Several existing visualization systems such as Many Eyes\cite{manyeyes},
Tableau\cite{tableau}, and Gapminder\cite{gapminder} support collaboration
asynchronously, but not in real-time. Though real-time remote collaboration is not a common feature to encounter in data visualization applications, we believe it has the potential to provide an engaging experience.

Existing systems that are most comparable to Weave include Instant
Atlas\cite{instant-atlas}, an ActionScript client that supports linked
visualizations with customizable layouts, and Choosel\cite{choosel}, a
visualization framework built using Google Web Toolkit\cite{gwt} and featuring a
windowing system that allows new visualizations to be dynamically created and
linked at runtime.

Developing a visualization system in ActionScript that can scale to reasonably
large data sets presents some challenges, since ActionScript is a
single-threaded programming language\cite{harui}, and the built-in event and
display list systems do not scale well.  These limitations have greatly
influenced the design of the Weave architecture.

This paper is structured to showcase the design of the Weave framework in
relation to other well-known, related frameworks.  The Weave framework is a
three-tiered design, separating the implementations of the sessioning, data, and
visualization layers. We begin by describing the core session framework in
detail, followed by brief descriptions of the data and visualization frameworks.

%%

% Results of probing and selection can be computed on the server as a remote
% procedure call.  Another option is for the client to download a simplified
% topology of the visualization that can be used to handle probing locally.  For
% example, it may degrade the response time of interactions such as probing and
% selecting  and in some situations these features aren't even supported. and
% provide apre-compute image tiles on the server and Images can be pre-computed
% and saved as tiles on a server to get the most responsive panning and zooming1

% Web-based visualizations come in many forms. There are systems that display
% plots as non-interactive, pre-computed images.

% WEAVE,
% OPENINDICATORS,
% GOALS OF WEAVE (SESSION HISTORY, COLLABORATION, )
% KNOWN BEFORE: (LAYERED VISUALIZATION, COLUMN BASED, BINARY REPRESENTATION=FAST)

% \section{Related work}
% UVP, LIBRARIES, (MANYEYES,
% \ldots

\section{Core Session Framework}
We describe Weave as a \emph{session-driven} application framework, which means
all significant actions made in the system are reflected in the session state,
and changes to the session state alter the state of the interface at runtime.
We define the session state of an object as the minimum amount of information
required to restore the current state of the object.

\begin{figure}[htbp] % h=here, t=top, b=bottom, p=page-of-its-own
 \centering
 \includegraphics[width=3.5in]{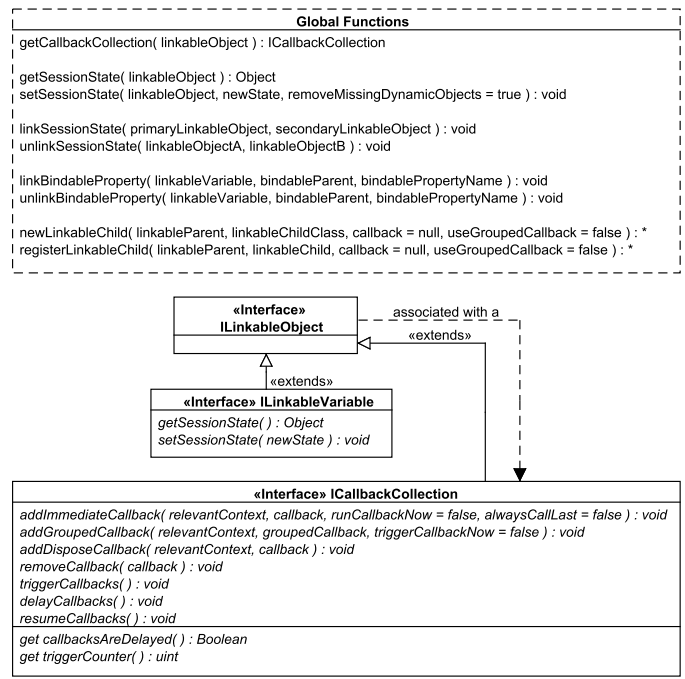}
 \caption{Core Weave interfaces. (Parameter types are omitted.)}
 \label{fig:core}
\end{figure}

The core theme in Weave code is to use linkable objects.  We use the term
\textit{linkable} to mean that an object has a mechanism for sending out signals
when a change occurs.  A linkable object may also have an associated session
state which may be defined explicitly or implicitly.  Figure \ref{fig:core}
shows some basic core interfaces in Weave related to linkable objects.

Each linkable object has a \textit{callback collection} associated with it,
which is used to signal changes in the linkable object and to manage a set of
functions to be called when a change is signalled.  If a linkable object is
registered as a child of another, the callbacks of the child will trigger the
callbacks of the parent.  This bubbling effect allows callbacks to be added to
higher-level objects that will be called when any descendant objects change.

\begin{figure}[htbp]
 \centering
 \includegraphics[width=3.5in]{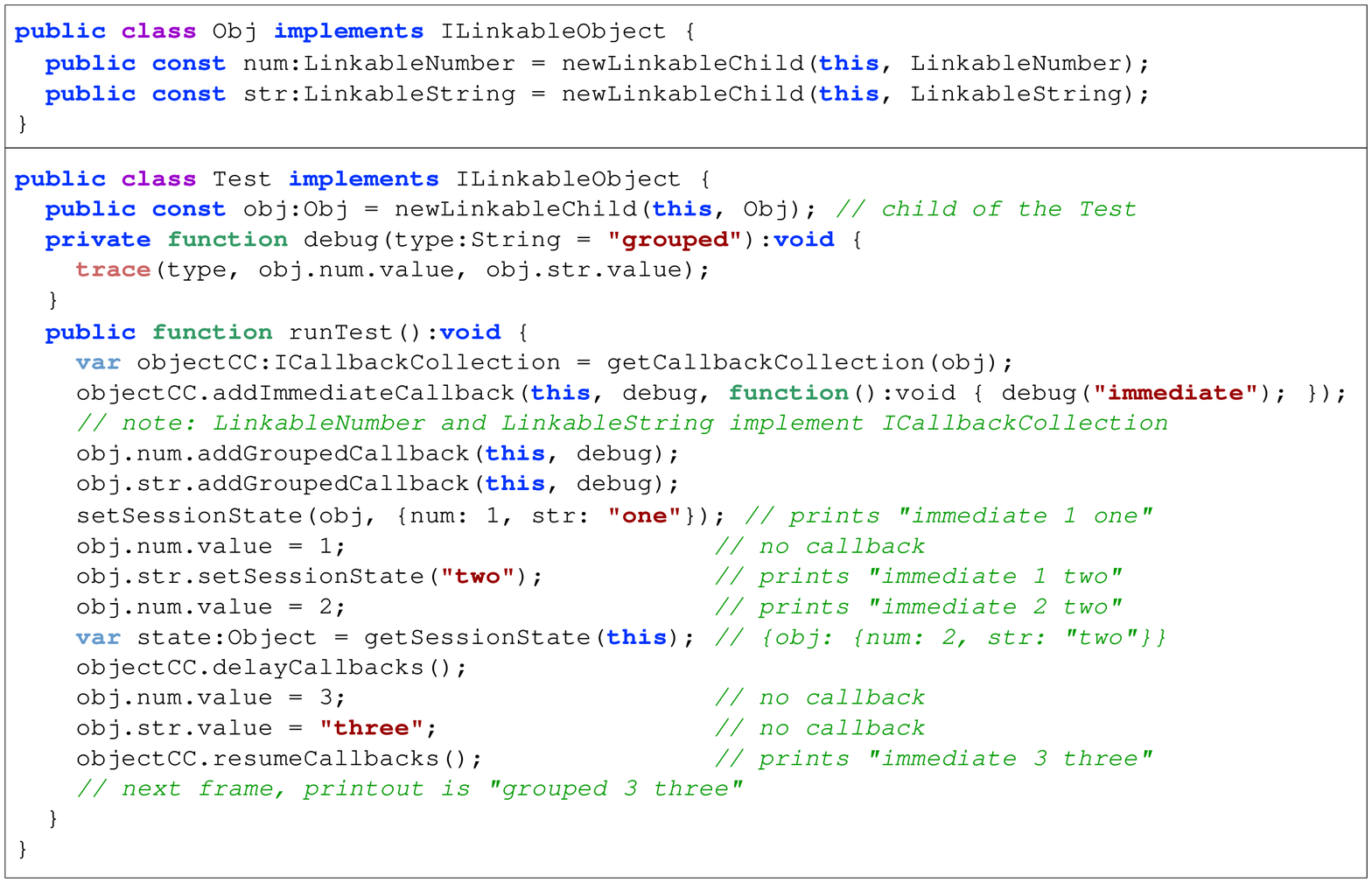}
 \caption{Sample code using linkable objects.}
 \label{fig:samplecode}
\end{figure}

The \textit{session state} for a linkable object can be defined explicitly by
implementing an interface, or implicitly by registering linkable child objects
that are publicly accessible as properties of the parent object.  To generate an implicit session state for
a linkable object in Weave, introspection\cite{introspection} is used to get a
list of public linkable property names.  Each property defining its session state
explicitly will be stored under the corresponding property name in a new dynamically
created object, while properties having implicit session states will be
recursively generated as nested objects. The resulting dynamic object is the
session state.  Figure \ref{fig:samplecode} provides sample code that
demonstrates callback collections and session states.

\subsection{Basic Linkable Objects}

The linkable object system implements a combination of the well-known observer
and memento design patterns\cite{gof}.  Implementations of these design patterns
can be found everywhere because they facilitate building complex reactive
systems with undo and redo capabilities.  We considered and implemented many
designs before coming up with the one presented in this paper.

\subsubsection{Observer Design Pattern}

\paragraph{ActionScript and Flex}
ActionScript provides its own implementation of the observer design pattern
called an \textit{event dispatcher}\cite{eventdispatcher}, which allows the
adding of event listeners that respond to dispatched event objects.  The Flex
SDK builds a feature on top of this event system called \textit{data
binding}\cite{binding}, which allows changes in a bindable property of an object
to be propagated.  This implementation has a few limitations that the Weave
framework addresses.
\begin{itemize}
  \item {\bf Event listeners must accept an event object as a parameter.} In
  practice, we have found that the information contained in event objects is
  rarely needed, since most of the relevant information can be gathered through
  other means.  This requirement is a nuisance and is what lead us to support
  callback functions with arbitrary parameter requirements in our
  \textit{addImmediateCallback} function shown in Figure \ref{fig:core}.
  \item {\bf Data binding uses string representations of property names.}
  This is required because a host object must exist in order to dispatch events
  when its primitive properties change.  This requirement causes maintainability
  issues because when you rename a bindable property, you also have to modify
  the strings used for binding to that property.  To avoid this issue, Weave
  does not use string representations of linkable property names, and instead
  uses object introspection to discover them.  Weave also eliminates the need
  for a host object by providing primitive classes such as LinkableString that
  implement the \textit{ILinkableVariable} interface, shown in Figure 1 and
  demonstrated in Figure 2.
  \item {\bf Data binding is one-way only.}\footnote{Flex 4 introduces
  two-way binding, but it is currently limited to MXML user interfaces
  and an ActionScript interface is not scheduled for implementation
  yet.\cite{2waybinding}} The data binding feature in Flex is useful for
  creating user interfaces that automatically update when internal variables
  change, but it is not suitable for two-way linking of arbitrary variables. 
  Weave provides the global \textit{linkSessionState} function, which
  creates a two-way linking of the session state between any two linkable
  objects, not just primitives.  The \textit{linkBindableProperty} function is
  also provided for a two-way linking between a linkable variable and a bindable
  property.
\end{itemize}
In contrast, ActionScript and Flex provide additional features in its event
system that are not supported by basic Weave interfaces.
\begin{itemize}
  \item {\bf Event listeners are associated with a specific type of event.}
  Weave makes no such distinction between changes in linkable objects.  This
  feature was not incorporated into Weave because for the most part, it was not
  needed.  In cases where it is needed, an additional callback interface is
  provided as a property of a linkable object that signals only a specific type
  of change.  This technique has proven to be sufficient for our purposes.
  \item {\bf Some event objects contain information about the event that cannot
  be gathered elsewhere.}  In Weave's basic callback interface, there is no
  place for information related to a specific change that triggers callbacks. 
  In the case where additional information is required, a secondary callback
  interface associated with a specific type of change provides the information
  relevant to the change that triggered the callbacks.  An example of this
  is shown later in Figure~\ref{fig:advanced}. This solves the issue without
  resorting to passing event objects around or requiring specific callback
  function signatures.
\end{itemize}

\paragraph{Push Versus Pull}
The observer design pattern described by Gamma et al. has two main parts: the
\emph{subject} and the \emph{observer}.\cite{gof} Our subject is the callback
collection, and our observers are callback functions. The observer described by
Gamma et al. is an object interface implementing a specific function signature.
The Java platform defines its observers in the same way, but adds an additional
argument comparable to the event parameter in ActionScript event
listeners.\cite{javaobserver} This is called the \emph{push} model, while Weave
mostly uses the \emph{pull} model, meaning observers must retrieve the
information themselves. However, inline functions can be created in ActionScript
to simulate the \emph{push} model as demonstrated in Figure~\ref{fig:samplecode}.
This flexibility allows the subject to remain ignorant about its observers and avoids
the issue of observers being dependent on a specific \emph{push} model implementation.

\paragraph{Spurious Updates}
One problem with the observer design pattern is unexpected or spurious
updates.\cite{gof}. Since each child linkable object in Weave triggers callbacks
of its parent and there is no distinction between different types of changes,
callbacks may be called unnecessarily and slow down the system.  Weave provides
a few features to help mitigate this problem.
\begin{itemize}
  \item {\bf Callbacks can be delayed.} The callback collection provides a way
  to delay callbacks so that multiple updates get grouped together.  This
  feature is used while setting the session state of a nested object to avoid
  running callbacks for every little change.  This feature is demonstrated in
  Figure \ref{fig:samplecode}.  Calling \emph{delayCallbacks} increases a
  counter, and \emph{resumeCallbacks} decreases it.  When the counter reaches
  zero, callbacks will be resumed.  This behavior allows nested function calls
  to delay and resume the same callback collection without running callbacks
  prematurely.
  \item {\bf Grouped callbacks treat multiple updates as a single one.}  While
  immediate callbacks may be called at any time, \emph{grouped callbacks} are
  only allowed to run during a scheduled time each frame. Callback collections
  use a central triggering system for grouped callbacks, which means that
  multiple updates will be grouped into a single one even if the updates come
  from different sources as demonstrated in Figure \ref{fig:samplecode}. 
  Because we cannot combine different sets of parameters into a single function
  call, grouped callbacks must require no parameters, as imposed by
  \emph{addGroupedCallback} shown in Figure \ref{fig:core}.
  \item {\bf Recursive triggering is disallowed on callback functions.} 
  Since callback functions may trigger other callback functions, infinite loops
  may occur if recursion is allowed.  In earlier versions of Weave, we provided
  a way to limit the recursion depth for each callback function.  The recursion
  limit did solve the problem, but we later discovered that specific control
  over the depth of recursion was not necessary, since the desired recursion
  depth was always zero when the feature was used.  The option for the
  recursion limit was changed to a boolean value and later removed because we
  found that recursion was not occurring where it was allowed, and it was never
  actually desired.
\end{itemize}

\subsubsection{Memento Design Pattern}
The purpose of the memento design pattern is to, ``without violating
encapsulation, capture and externalize an object's internal state so that the
object can be restored to this state later.''\cite{gof}  Adopting this design
pattern allows advanced features to be developed, such as session history with
undo/redo and real-time collaboration.  However, ease of development depends
greatly on the details of the implementation.

\paragraph{Encapsulation Versus Simplicity} Encapsulation is difficult to
achieve in ActionScript without a severe performance hit. Since the \emph{const}
qualifier in ActionScript lacks the expressiveness of the \emph{const} qualifier
in C++, complex private member variables such as Arrays can only be fully
protected by returning a copy of the object.  Weave is designed to be scalable
and in most situations encapsulation is a secondary concern, so we avoid making
copies of objects in many situations even though it violates encapsulation.  The
session state generated from a linkable object consists of primitive immutable
types, so generating the session state does not cause an encapsulation problem. 
However, the linkable children that define the implicit session state of an
object are publicly accessible. This design was chosen for simplicity. The
example in Figure \ref{fig:samplecode} demonstrates how simple it is to define
the session state of a linkable object, requiring only one line of code per
child object.

\paragraph{Explicit Versus Implicit} The memento implementation suggested by
Gamma et al. requires explicitly defining the session state for each object you
want to save and restore.  This is done by defining a separate memento class for
each type of object.  This method is used by the Swing framework\cite{swing} in
Java and by Lott et al.\cite{as-design-patterns} in ActionScript.  The Swing
framework does automatically include nested GUI components in the session state
of an application, but the session state for each component is explicitly
defined.  We feel that creating a separate memento class for each object is
cumbersome and instead prefer implicitly generated session states. Our first
attempt at the memento design pattern did involve explicitly defining the
session state for a limited number of objects, but as the demands for
flexibility grew, we realized that we needed to automatically generate session
states for developer efficiency.

\paragraph{Object Serialization} Object serialization is another form of the
memento design pattern.  To serialize an object is to write its session state to
a stream so that it can be recreated elsewhere.  Many platforms allow object
serialization, including Java\cite{java-serialize},
ActionScript\cite{as-bytearray}, and Microsoft .NET\cite{dotnet}.  Like Weave,
Java and .NET allow the session state of an object to be defined implicitly or
explicitly.  However, while Weave requires that you specify which objects should
be included in the session state with the empty \emph{ILinkableObject}
interface, Java and .NET require that you specify which objects you want to
skip.  Our reasoning is that there will be much fewer objects required to
restore the session state than those that are not, since many properties are
derived from others. For example, when extending complex GUI components in
ActionScript that we cannot modify, there will be many properties that are
irrelevant for typical usage.\footnote{Although display objects cannot be
serialized in ActionScript, this is still a valid point.} This technique also
avoids including unnecessary information when real-time collaboration is
implemented. Session states in Weave change often, and it would not make sense
to always create new objects when the state changes, so object deserialization
is done by modifying the state of an existing object. Java uses this same
technique\cite{java-serialize}, but deserialization in
ActionScript\cite{as-bytearray} and the .NET framework\cite{dotnet} always
creates a new object.

\paragraph{Incremental Updates} Regarding support for restoring session state, a
related method is to implement undoable commands in the command design
pattern\cite{gof}.  This method is used by related visualization applications
such as the recent Choosel framework\cite{choosel}. Weave is more oriented
towards saving and restoring the state of the entire application rather than
undoing individual commands, so we do not implement this design pattern.  A
benefit of the command pattern is that it supports incremental changes suitable
for undo and redo operations. However, if a system implements session history
completely with undoable commands, it may make navigating to distant session states
inefficient compared to restoring a complete snapshot of the application state. 
Weave allows both the restoring of entire application states and incremental
changes by allowing partial session states to be restored. Support for this feature
involves setting the \emph{removeMissingDynamicObjects} parameter to false in
the global \emph{setSessionState} function shown in Figure \ref{fig:core}.

\subsection{Advanced Linkable Objects}
\begin{figure}[htbp] % h=here, t=top, b=bottom, p=page-of-its-own
 \centering
 \includegraphics[width=3.5in]{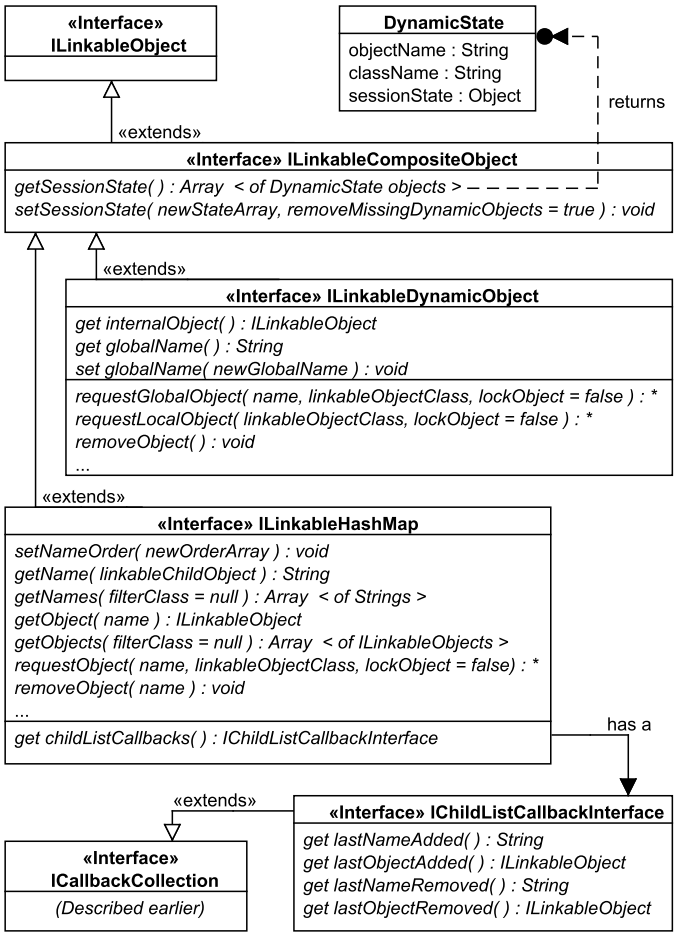}
 \caption{Advanced Weave interfaces. (Some details are omitted.)}
 \label{fig:advanced}
\end{figure}

To support a complex windowing system suitable for collaborative data
visualization and exploration, we must have a way to dynamically create and
refer to objects from a session state.  For this purpose we introduce the
\emph{DynamicState} object, shown in Figure \ref{fig:advanced}, which contains
the required information.  To make use of this structure, we define the
\emph{ILinkableCompositeObject} interface for an object having a session state
explicitly defined as a list of DynamicState objects.  Defining the session
state this way allows any number of linkable objects to be dynamically created
and destroyed at runtime through the session state interface.  Figure
\ref{fig:advanced} shows two key interfaces in Weave that extend
ILinkableCompositeObject.

\subsubsection{Linkable Hash Map}
\begin{figure}[htbp] % h=here, t=top, b=bottom, p=page-of-its-own
 \centering$
    \begin{array}{cc}
    \includegraphics[width=1.14in]{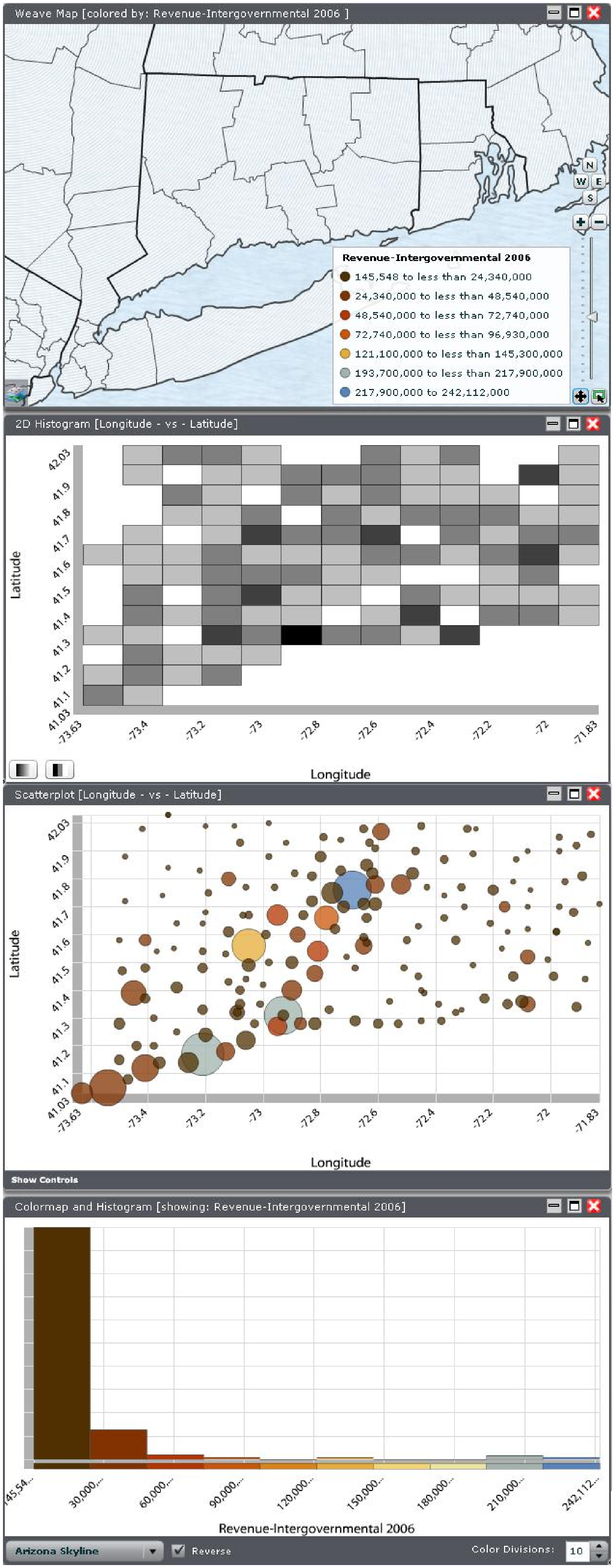}
    \includegraphics[width=0.03in]{blank.eps}
    \includegraphics[width=1.14in]{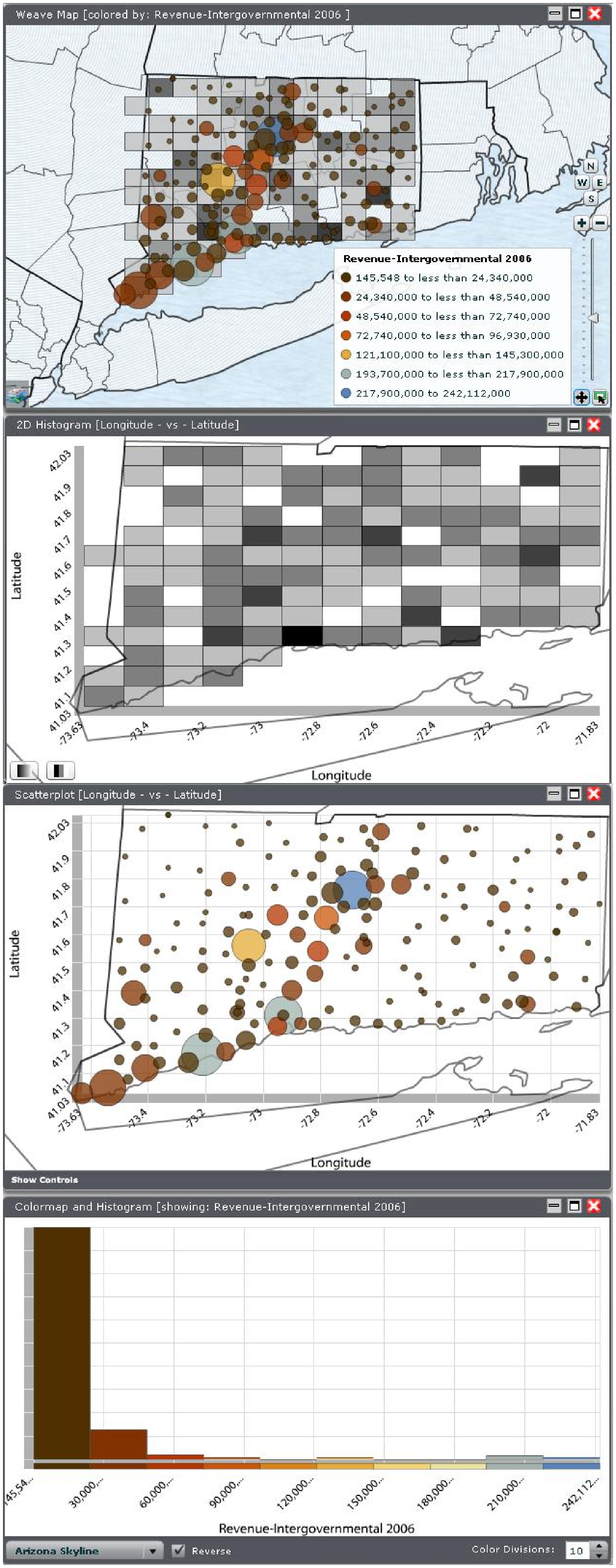}
    \includegraphics[width=0.03in]{blank.eps}
    \includegraphics[width=1.14in]{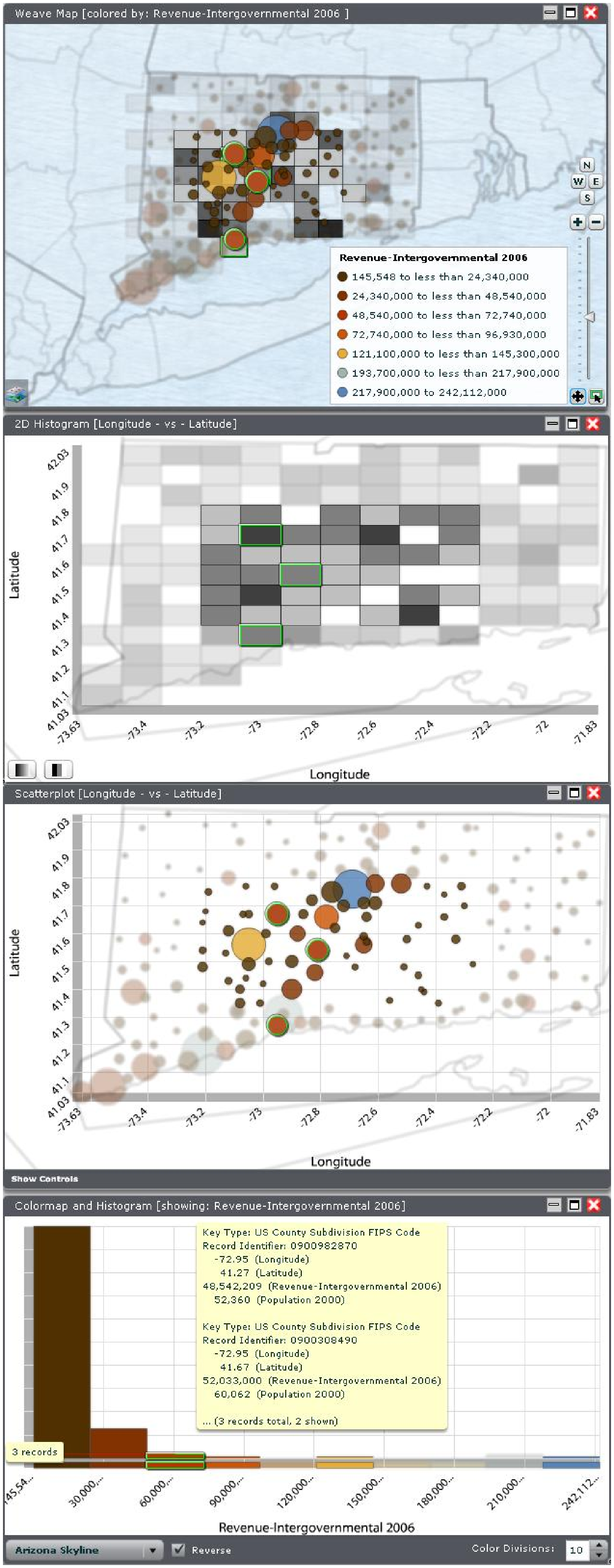} \\
    \includegraphics[width=2.4in]{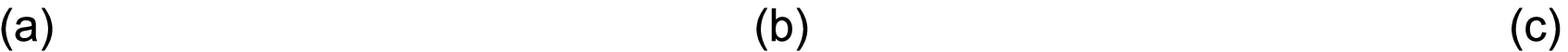}
    \end{array}$
 \caption{Examples of advanced session state capabilities: (a) Sample Weave
 visualizations. (b) Mashed-up visualizations. (c) Linked selection and
 probing.}
 \label{fig:mashup}
\end{figure}

\emph{ILinkableHashMap} is an interface to an ordered list of named child
objects that can be dynamically created, destroyed, and reordered at runtime.
The only requirement for a dynamically created object is that it implements
ILinkableObject so that its state can be saved, restored, and
monitored.\footnote{We are currently experimenting with dynamic creation of any
non-linkable object by defining the session state as a dynamic subset of its
public properties.} For the life of a dynamically created object, its name
remains the same while the order is allowed to change.  This behavior is ideal
for a windowing system so the window handles remain the same while the
z-ordering may change.  The order of objects can also be used for other purposes
such as the z-ordering of visualization layers and the order of dimensions in a
stacked bar chart or parallel coordinates plot. Figure \ref{fig:mashup}(a) and
(b) illustrates the capability of adding new visualization layers dynamically
using an ILinkableHashMap.

The root object in a Weave session state implements ILinkableHashMap to act as a
blank slate that may take on any behavior at runtime.  To make dynamically
created display objects appear on the screen, a simple utility function is
provided in Weave that synchronizes a GUI container with the session state of an
ILinkableHashMap.  With the addition of a plug-in system that
allows runtime linking of libraries, Weave gains flexibility comparable to the
WebCharts\cite{webcharts} visualization framework. Moreover, since the core session framework is
not limited to visualization tools, the plug-in and collaboration
systems for Weave open up a much wider range of capabilities.

\subsubsection{Linkable Dynamic Object}
\emph{ILinkableDynamicObject} defines an interface to a wrapper for a single
dynamically created linkable object.  An implementation of this interface can
adapted for use in numerous design patterns described by Gamma et al. that
involve runtime swapping and wrapping of objects.  Weave uses this interface to
allow runtime swapping of visualization plotters, data sources, and binning
definitions.

Besides being a placeholder for a dynamically created object, the
ILinkableDynamicObject interface allows referencing a global object by name.
\emph{Global linkable objects} reside in the root ILinkableHashMap of Weave, and
the global names that identify them are the same names used in the hash map.
Global object references allow the session state to define dynamic linking of
visualization properties such as color mappings, record subsets, record
selections, actively probed records, and visualization attributes. Figure
\ref{fig:mashup}(c) illustrates linking of color, selection, and probing across
visualizations in Weave.

\section{Data Framework}
Currently, Weave supports relational data from CSV files, XLS files, DBF files,
WFS services, and a custom server that provides SQL data. The only requirement
Weave has for importing data from any of these sources is that each record has a
primary key (unique record identifier) associated with it. The reason for this
is that we want Weave to be able to generate visualizations containing data from
multiple sources.  The keys provide a way to match up records from different
sources that may have missing records or provide them in a different order.
However, we cannot use keys alone to match up records.  We also need a key type
(a namespace) to qualify the keys.  Weave uses qualified keys as globally
unique record identifiers for associating records from different columns. The
key type system in Weave is mainly used to prevent users from mistakenly
associating unrelated data to produce meaningless visualizations.  For example,
if a geographic map visualization has a layer for towns and a layer for schools
which happen to have similar key values, we don't want to display town-level
data on the school layer, and vice versa.

The key typing system in Weave currently allows key types to be arbitrary
strings. In future versions, we would like to define a standard set of key types
identified by URIs, such as
\emph{\url{http://www.openindicators.org/keyTypes/US-State-FIPS-Code}}, and
provide a service for selecting from that standard set of key types when
importing data into Weave.  If multiple sites begin using standard, well-defined
key types when publishing their data through Weave servers, compatible data from
these sites can be matched up automatically.

As an extension of the key type concept, we would also like to define or
reference a standard set of data units in the same way. With unit information,
automatic conversions can be made such as feet to meters or dollars to euros. 
Another use for the unit property would be to tell Weave that a particular
column of data refers to a foreign key in another data set, which would allow
for automatic aggregation of data in hierarchical data sets.

Because the Weave client supports data coming from servers with varying
implementations, automatic record mapping and unit conversions must be
implemented in the client side.  The implementation will be generalized to
support metadata about columns of data from any type of data source.  A
server-side component would also be created to store this column-level metadata.

% - equations and data transformations
%Weave has a generalized data structure for storing columns of data.

\section{Visualization Framework}
%\subsection{Single-Threaded Environment}
The ActionScript virtual machine is single-threaded.  This means that during
heavy computation, the interface will become unresponsive\cite{harui}.  Because
Weave wants to process large amounts of data, this limitation had to be kept in
mind as the framework was implemented.  Three main bottlenecks were encountered
during devleopment: text and data processing, rendering overlapping vector
graphics, and garbage-collection.  These three activities were found to
contribute the most to an unresponsive interface.

% \subsection{Avoiding Unnecessary Text Processing}

% \begin{itemize}
% \item FIRST TEST: XML ARRAY REPRESENTATION (ALSO COMPARE JSON)
% \item EARLY WEAVE: WEBSERVICES WITH BASE64 ENCODING
% \item WEAVE 1.0: AMF3 ENCODING (Sri)
% \end{itemize}

% \paragraph{Overcoming Graphics Engine Limitations}
The natural way to implement a data visualization system in ActionScript is to
use display objects for each data record or shape, such as in Flex Charting
\cite{flexcharting} or Flare \cite{flare}. However, when thousands of objects
are added to the display list a typical application will slow down dramatically.
This was a known issue at the beginning of the Weave project, so instead of
adding objects to the display list, the objects were kept off-screen and
associated vector graphics were drawn directly to a visualization layer when
they were needed. This method resulted in decent performance for a few thousand
records, but not for 10,000 and above. We were not satisfied with this
limitation, so we sought to determine the most scalable rendering method.

It was discovered that not only does a long display list slow down the
application, but overlapping vector graphics was slower to render compared to
non-overlapping graphics.  This behavior was verified by randomly generating
10,000 circles on a canvas with random colors. When the range of random numbers
was constrained more, the rendering became slower. The slowest performance
occurred when all the circles were drawn at the same location.  This means that
in a typical scatterplot implementation, the time required to render would
depend on the spread of the data on the screen.  The fastest rendering method
for an equivalent plot of circles turned out to be using the \textit{copyPixels}
function with \textit{BitmapData} objects.  To avoid the issues with overlapping
vector graphics and allow for the possibility of fully optimized bitmap
implementations, Weave was refactored to render plot graphics directly to
bitmaps.

% \subsection{Avoiding Garbage Collection}
% % Throughout the Flex framework, new objects are created often.  Every
% % event that gets dispatched is a new object.  Whenever you convert a Point object from local
% % to global coordinates, you get back a new object.
% % 
% % A standard approach for avoiding garbage-collection is to use object pooling.
% \begin{itemize}
% \item TYPICAL: DISPLAY OBJECT FOR EACH DATA POINT
% \item EARLY WEAVE: SAVE OFF-SCREEN OBJECTS FOR EACH DATA POINT, STILL TOO SLOW
% AND MEMORY INTENSIVE (DUPLICATION OF DATA)
% \item WEAVE 1.0: NOW USES ONLY SPATIAL INDICES (REQUIRED FOR SELECTION/PROBING)
% THAT USE OBJECT POOLING
% \end{itemize}

\section{Conclusion}
We have developed a foundation for a general web-based application framework
with broad, expanding goals in mind to enable data visualization, exploration,
analysis, session history, and real-time collaboration. We have described the
core Weave framework in detail, emphasizing ideals such as simplicity,
maintainability, and flexibility. We have related our design decisions to those
of other well-known systems and provided the reasoning behind our decisions.

Weave is the result of two years of development and we are now planning years
three through five.  Now that the initial framework is in place, our specific
research goals are to further develop the framework into a platform that can
support a variety of future research topics at the Institute for Visualization
and Perception Research.  Weave is open source software released under MPL 2.0 and available at \url{http://github.com/adufilie/Weave}.

% We have demonstrated that our framework allows rapid development of applications
% driven by changes in session state, and introduced the possibilities that our
% framework opens up.

% We believe the collaboration of the OIC has pushed Weave to
% become much general and flexible than any of us had expected.

%- sessioning framework allows rapid prototyping of new plotters during runtime

% % if specified like this the section will be ommitted in review mode
\acknowledgments{ The Weave project is funded by members of the OIC, the Knight
Foundation\cite{knight} and the University of Massachusetts at Lowell. Weave is
supported by a twenty-member software and data development team led by
University of Massachusetts faculty members, Dr. Georges Grinstein and Dr.
William Mass and involving numerous Computer and Social Science masters and
doctoral students. The design of the current Weave framework and its further
evolution to support future research topics constitute the core of a Ph.D.
thesis to be carried out by Andrew Dufilie at the University of Massachusetts
Lowell under his advisor, Georges Grinstein. }


\begin{thebibliography}{50}

\bibitem{oic} "Open Indicators Consortium (Weave)." National Neighborhood Indicators Partnership. (\url{http://www.neighborhoodindicators.org/activities/projects/open-indicators-consortium-weave}).

\bibitem{baumann} Baumann, A., Grinstein, G., and Mass, W.: Collaborative Visual Analytics with Session Histories. Unpublished Research Paper. No. 2009-005, University of Massachusetts Lowell, Dept. of Computer Science, Lowell, MA 01854. (\url{http://teaching.cs.uml.edu/~heines/techrpts/details.jsp?Year=2009&SeqNo=005}).

\bibitem{tableau} Tableau Public. (\url{http://www.tableausoftware.com/public}).

\bibitem{gapminder} Gapminder. (\url{http://www.gapminder.org}).

\bibitem{manyeyes} Many Eyes. (\url{http://www-958.ibm.com/software/data/cognos/manyeyes})

\bibitem{ubiquity} "Flash Player Version Penetration." December, 2010. (\url{http://www.adobe.com/products/player_census/flashplayer/version_penetration.html})

\bibitem{instant-atlas} Instant Atlas. (\url{http://www.instantatlas.com}).

\bibitem{choosel} L. Grammel and M.-A. Storey. "Choosel - Web-based Visualization Construction and Coordination for Information Visualization Novices." In InfoVis '10: Proceedings of the IEEE Symposium on Information Visualization, 2010.

\bibitem{gwt} "Google Web Toolkit." Google Code. Accessed 2011. (\url{http://code.google.com/webtoolkit/}).

\bibitem{harui} A. Harui. "Threads in Actionscript 3." Weblog Entry. Alex's Flex Closet. January 1, 2008. (\url{http://blogs.adobe.com/aharui/2008/01/threads_in_actionscript_3.html}).

\bibitem{introspection} "Performing object introspection." Adobe Flex 3 Help. 2008. (\url{http://livedocs.adobe.com/flex/3/html/help.html?content=usingas_8.html}).

\bibitem{gof} E. Gamma, R. Helm, R. Johnson, and J. Vlissides. "Design Patterns." Addison-Wesley, Boston, MA, 1995.

\bibitem{eventdispatcher} "IEventDispatcher." ActionScript 3.0 Language and Components Reference. 2008. (\url{http://livedocs.adobe.com/flash/9.0/ActionScriptLangRefV3/flash/events/IEventDispatcher.html}).

\bibitem{binding} "Using the Bindable metadata tag." Adobe Flex 3 Help. 2008. (\url{http://livedocs.adobe.com/flex/3/html/help.html?content=databinding_8.html}).

\bibitem{2waybinding} "Two-way Data Binding - Functional and Design Specification." WikiEntry. Adobe Open Source. 2009. (\url{http://opensource.adobe.com/wiki/display/flexsdk/Two-way+Data+Binding}).

\bibitem{javaobserver} "Observer." Java 2 Platform SE 6. (\url{http://download.oracle.com/javase/6/docs/api/java/util/Observer.html}).

\bibitem{swing} "What is Swing?" The Java Tutorials. Accessed 2011. (\url{http://download.oracle.com/javase/tutorial/ui/overview/intro.html}).

\bibitem{as-design-patterns} J. Lott and D. Patterson. "Advanced ActionScript 3 with Design Patterns." Peachpit Press, Berkeley, CA, 2007.

\bibitem{java-serialize} "Serializable." Java 2 Platform SE 6. (\url{http://download.oracle.com/javase/6/docs/api/java/io/Serializable.html}).

\bibitem{as-bytearray} "Reading and writing a ByteArray." Adobe Flex 3 Documentation. 2008. (\url{http://livedocs.adobe.com/flex/3/html/help.html?content=ByteArrays_2.html#1047174}).

\bibitem{dotnet} P. Obermeyer and J. Hawkins. "Object Serialization in the .NET Frame-work." MSDN. (\url{http://msdn.microsoft.com/en-us/library/ms973893.aspx}).

\bibitem{webcharts} D. Fisher, S. M. Drucker, R. Fernandez, and S. Ruble. "Visualizations Everywhere: A Multiplatform Infrastructure for Linked Visualizations" InInfoVis '10: Proceedings of the IEEE Symposium on Information Visualization, 2010.

\bibitem{flexcharting} "About charting." Adobe Flex 3 Help. 2008. (\url{http://livedocs.adobe.com/flex/3/html/help.html?content=Part1_charting_1.html}).

\bibitem{flare} UC Berkeley Visualization Lab: Flare Data Visualization for the Web. Tutorial. 2009. February 13, 2011. (\url{http://flare.prefuse.org/tutorial}).

\bibitem{knight} Fest, M.: Knight Foundation Spurs New Round of Local News and Information Projects Nationwide. Press release. Knight Foundation. January 13, 2010. (\url{http://www.knightfoundation.org/news/press_room/knight_press_releases/detail.dot?id=355583}).

\end{thebibliography}
\end{document}